 \documentclass[manuscript]{aastex}

\usepackage{emulateapj5}

\def\bh{black hole~}
\def\bhs{black holes~}
\def\el {emission-line~}
\def\els {emission-lines~}
\def\ed {Eddington~}
\def\ers{{\rm erg/sec}}
\def\ms{M_{\odot}}
\def\et{et al.\ }
\def\rev{reverberation~}
\def\vFWHM{\ifmmode v_{\mbox{\tiny FWHM}} \else
            $v_{\mbox{\tiny FWHM}}$\fi}
\def\kms{\ifmmode {\rm km\ s}^{-1} \else km s$^{-1}$\fi}
\def\ers{\ifmmode {\rm erg\ s}^{-1} \else erg s$^{-1}$\fi}

\def\aa{{\it Astronomy and Astrophysics}}

\newenvironment{inlinefigure}{%
\def\@captype{figure}%
\noindent\begin{minipage}{0.999\linewidth}\begin{center}}         
{\end{center}\end{minipage}\smallskip}

\slugcomment{The Astrophysical Journal, submitted July 11, accepted October 12}

\shorttitle{The Black Hole-Bulge Relation Revisited}
\shortauthors{Wandel}

\begin{document}

\title{Black Holes of Active and Quiescent Galaxies: \\
I. The Black Hole-Bulge Relation revisited}

\author{Amri Wandel}

\affil{Racah Inst. of Physics, The Hebrew University, Jerusalem 91405,Israel}
\email{amri@vms.huji.ac.il}

\begin{abstract}
Massive Black Holes detected in the centers of many nearby galaxies show an approximately linear relation
with the luminosity of the host bulge, with 
the black hole mass being 0.001-0.002 of the bulge mass. Previous work suggested that 
black holes of  active (Seyfert 1) galaxies follow a similar relation, but apparently with a  
significantly lower value of $M_{\rm BH}/M_{\rm bulge}$ (Wandel 1999).
New data show that 
this difference was mainly due to over-estimating the black hole mass in quiescent
galaxies and over-estimating the bulge magnitude of Seyfert galaxies.
Using new and updated data we show that 
AGNs (Seyfert galaxies and quasars) follow the same BH-bulge relation as ordinary (inactive) 
galaxies. We derive the BH/bulge relation for a sample of 
55 AGNs and 35 quiescent galaxies, finding
that broad line AGNs have an average black hole/bulge mass fraction of $\sim 0.0015$ with a strong correlation ($M_{\rm BH}\propto L_{\rm bulge}^{0.9\pm 0.16}$). This BH-bulge relation is consistent with the BH-bulge relation of quiescent galaxies and much tighter than previous results.
Narrow line AGNs appear to have a lower ratio, $M_{\rm BH}/M_{\rm bulge}\sim 10^{-4}-10^{-3}$.
We find this to be a more general feature, the BH/bulge ratio in AGNs being 
inversely correlated with the emission-line width, implying
a strong linear relation between the size of the broad emission line region and the luminosity of the bulge. Finally,
combining AGNs with observed and estimated stellar velocity dispersion, we find 
a significant correlation ($M_{\rm BH}\propto \sigma ^{3.5-5})$, consistent with 
that of quiescent galaxies.
 
\end{abstract}


\section{Introduction}

Compact dark masses, probably massive \bhs (MBHs), have been detected in the cores of 
many nearby galaxies  (Kormendy and Richstone 1995, Kormendy \& Gebhardt (2001)).
Magorrian \et (1998) presented a sample of 32 nearby galaxies with MBHs 
and suggested that the \bh (BH) mass is proportional to the luminosity of the host galaxy or host
bulge (or equivalently, to their mass) hereafter referred to as the BH/bulge relation,
with the BH mass being about 0.006 of the mass of the spheroidal bulge (although with a 
significant scatter). 

In addition to the MBHs detected by techniques of stellar and gas kinematics,
the masses of about three dozen  MBHs in AGNs have been estimated  by reverberation 
mapping of the broad emission-line region. 
High quality reverberation data and virial BH mass estimates are presently available 
for 20 Seyfert 1 nuclei (Wandel, Peterson and Malkan 1999, hereafter WPM) 
and 17 PG quasars (Kaspi \et 2000).
Laor (1998) suggested that  the masses of BHs in quasars (estimated 
empirically) and their host bulges follow the BH/bulge relation of Magorrian \et (1998).
Using the WPM Seyfert sample with bulge estimates from Whittle (1992) Wandel 
(1999) found that the BH/bulge ratio is significantly lower (on average, by a 
factor of $\sim$20) than the ratio found by  Magorrian  \et (1998) for 
MBHs in quiescent galaxies. 

Recent research using higher quality HST data and a more careful treatment of the
modeling uncertainties give lower values for \bh masses in nearby galaxies
(Merritt and Ferrarese 2001a),  with an average \bh to bulge mass ratio of $\sim$0.001  
and a nearly linear BH-bulge relation
 $M_{\rm BH}\propto L_{\rm bulge}^{1.1}$ (Kormendy \& Gebhardt 2001).

The updated lower dynamical BH mass estimates for quiescent galaxies reduced the
discrepancy between inactive galaxies and \rev mapped Seyferts, but
Consequently,  the quasars in 
Laor's (1998) sample appear to have larger BH-bulge ratios than inactive galaxies and of less luminous AGNs.
To explain this Laor (2001) suggested a nonlinear BH-bulge mass relation, $M_{\rm BH}\sim M_{\rm bulge}^{1.5}$.

Ferrarese and Merritt (2000) and Gebhardt \et (2000a) have found that the mass of MBHs in quiescent
galaxies is better correlated with the stellar velocity dispersion in the bulge 
than with the bulge luminosity.
Although to date there are only a few (less than ten) Seyfert galaxies with reliable measurements of stellar velocity and with reverberation BH mass
estimates (Ferrarese \et 2001), they seem to be consistent with the BH-velocity dispersion relation of inactive galaxies. 

Recently McLure \& Dunlop (2001, hereafter MD01) have studied the BH-bulge relation of 30 quasars, 
combined with the
Seyfert galaxies in the WPM sample for which they estimate bulge magnitudes from  HST imaging and 
bulge-disk decomposition. Assuming a flattened BLR geometry and a biased distribution of 
inclinations MD01 suggest that the actual BH mass of AGNs is 3 times larger than inferred from the 
reverberation virial method assuming isotropic geometry. 
While at present a factor of 2-3 in the BH 
reverberation mass cannot be 
ruled out, systematically increasing the Seyfert BH masses by a factor of 3 seems to violate the agreement between
Seyferts and quiescent galaxies in the BH-velocity dispersion relation (fig. 2 of Ferrarese \et 2001).

In this work we reconcile the BH-bulge mass relation of AGN with that of quiescent galaxies, analyzing the 
factors which led to the earlier apparent discrepancies between Seyferts, quasars and inactive galaxies, and
demonstrate that all three groups do follow  the same 
BH-bulge mass relation without a spatial geometry or bias for AGNs.
In section 2 we present the
sample, including 23 Seyfert 
galaxies, 32 quasars and 35 inactive galaxies with reliable $M_{\rm BH}$ and  bulge data.  
In section 3 we show that the 
BH- bulge mass relation of the combined Seyfert+quasar sample is consistent with 
that of inactive galaxies, both in slope and in normalization,
discuss the uncertainties and point out the factors that produced the apparent 
discrepancy in the $M_{\rm BH}-M_{\rm bulge}$ relation between Seyferts and inactive galaxies 
in previous works.
We also find that Narrow Line Seyfert 1s and narrow line quasars have significantly lower 
$M_{\rm BH}/M_{\rm bulge}$ ratios than broad line AGNs and inactive galaxies,
which is the result of a systematic trend (section 4) - a 
strong correlation between the BH mass to bulge luminosity (or mass) ratio and
the emission line width of the active nucleus.
In section 5 we analyse the implications of the BH mass - stellar velocity dispersion relation, 
while section 6 discusses possible reasons for the lower BH/bulge ratio of narrow line AGNs. 

\section{Data}

We compile a sample of 55 AGNs with BH mass and bulge magnitude. 
The AGN data are listed in tables 1 (Seyferts) and 2 (quasars). We use 
$H_o=80\kms{\rm Mpc}^{-1}$ and $q_0=0.5$.
We convert magnitudes from other works to the V band, using the relations $V-R=0.8$ and $B-V=0.8$. 
We express the bulge magnitude in units of $L_{\odot}$ using the standard relation
$L_{\rm bulge}/L_\odot=0.4(-M_V+4.83)$.
To find the bulge mass we 
use the relation
$M/M_\odot=1.18(L/L_\odot)-1.11$
(Magorrian \et 1998). 

\subsection {Seyferts}
We use the sample of 17 Seyferts in Wandel, Peterson and Malkan (1999), 
supplemented by 6 more Seyferts with reliable BH mass or bulge magnitude data:
MRK 279 (Santos Lleo \et 2001), NGC 3516 and
NGC 4593 (Ho 1999, Gebhardt \et 2000b) and
MRK 841, NGC 4253 and NGC 6814 (bulge magnitudes from Virani \et , BH mass from L-R relation 
(eq. 2 below, cf. Laor 2001). For the Seyfert galaxies in the 
WPM sample we use the bulge magnitude given by MD01, 
estimated from HST imaging and two dimensional 
bulge-disk decomposition. This method is more accurate than 
the empirical formula for the bulge/total ratio depending on the Hubble type 
(Simien \& deVaucouleurs 1986;  Whittle \et 1992) used in Wandel (1999).
The bulge luminosity obtained by bulge/disk decomposition of galaxy images is systematically lower than the empirical estimate: for the 15 
Seyferts common to Wandel (1999) and MD01 the latter have an average bulge 
magnitude lower by a factor of 3. 
However, the difference in the bulge magnitude estimated by the two methods
is not uniform: for individual objects it varies between 1 and 10. We find that the 
correction factor (that is, the ratio of the bulge luminosity derived by MD01 from HST imaging and that 
estimated from the Simien \& deVaucouleurs (SdV) relation) is correlated with the FWHM of the broad
emission lines (fig. 1). For example, while the Seyferts with the broadest lines 
(F9, IC 4329A, NGC 3227 and NGC 4151) have 
correction factors of order 10, the narrow line objects (MRK 335 and the marginally narrow line 3C120) have a negligible correction,
$\approx 1$. 
The best fit to the bulge luminosity correction  for 
the 15 Seyferts in MD01 is 
$$\log (L_{\rm SdV}/ L_{\rm img})=1.3 \log \left ( {{\rm FWHM(H}\beta)\over
10^3\kms}\right ) -0.17  \eqno (1)$$
The correlation  coefficient is R=0.77 with a  $<0.1\%$ probability of chance correlation. This empirical correlation is so tight that we use it to predict the actual bulge magnitude from the SdV relation
for Seyferts which do not have directly measured bulge luminosity. 
With the empirical formula we obtain bulge magnitudes
for the five Seyferts in our sample which do not have measured bulge luminosity (AKN 120, MRK 110, MRK 279, 
NGC 3516, NGC 4593).
In the above analyses we have used for 3C390.3 a FWHM H$\beta$=3500\kms (Crenshaw 1986) 
which is substantially lower than the
value of $10^4\kms$ reported in WPM; this very broad feature may be temporary, due to the line profile 
variability (Peterson, private communication). 

\begin{inlinefigure}
\centerline{\includegraphics[width=1.0\linewidth]{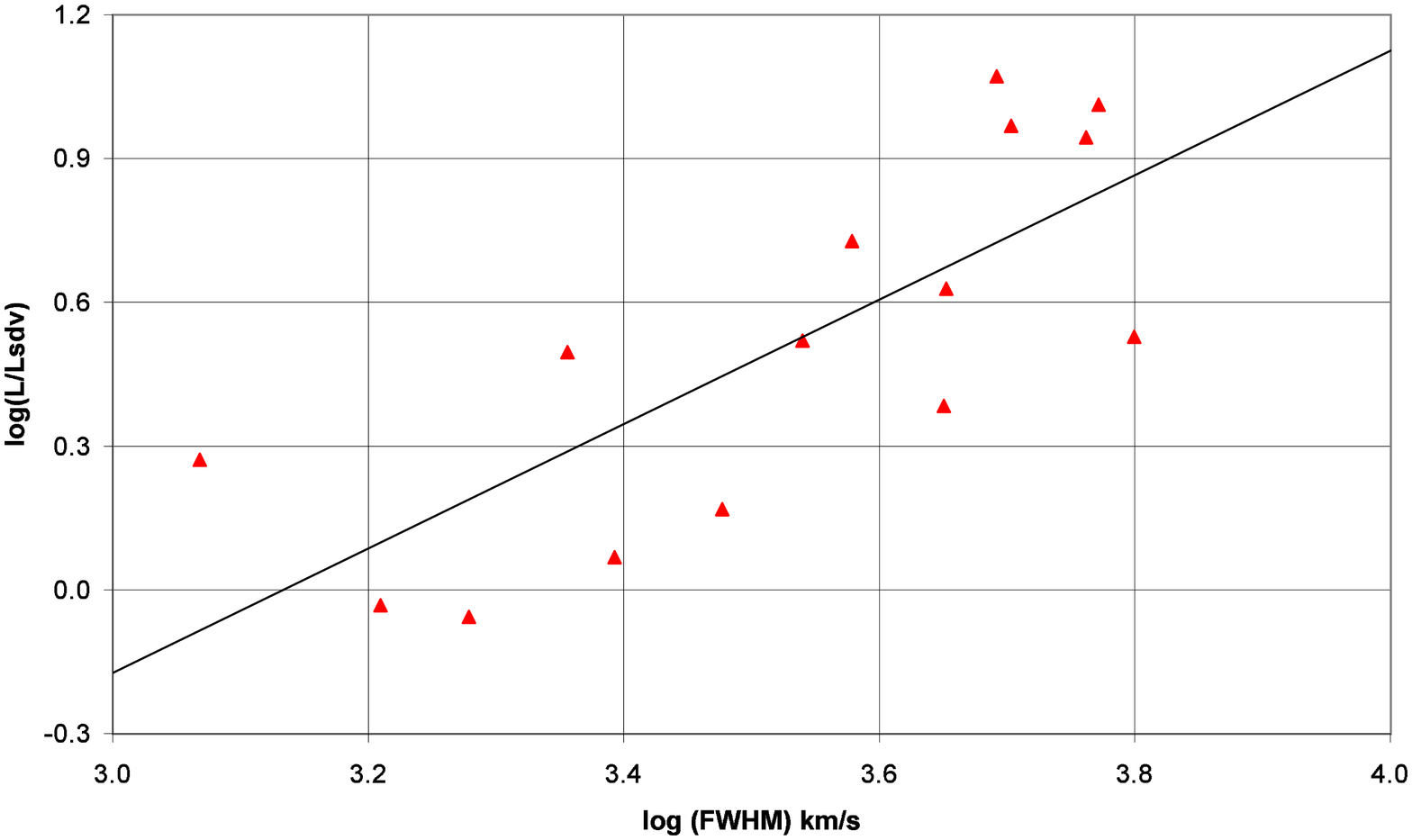}}
\figcaption{ The bulge luminosity correction: the difference between the bulge/disc decomposition method (MD01) 
and the empirical Simien-de Vaucouleurs method (bulge/total dependence on
Hubble type) vs. the H$\beta$ line width for Seyfert 1 galaxies.
}
\end{inlinefigure}

\subsection{Quasars}
We adopt the quasar sample of MD01, supplemented by PG1425+267 (Kirhakos \et 1999) and PG1704+60 
(BH mass from Kaspi \et 2000, host magnitude from Hamilton, Casetano and Turnshek 2000).

\begin{table*}
\vspace{-0.7cm}
\begin{center}%
\vspace{-0.5cm}
\caption{Seyfert data}
\scriptsize
\begin{tabular}{lrrrrrrrr}
\tableline\tableline
{}&{}&{}&{}&{}&{}&{}&{}&{}\\
Name   &  FWHM &   log$L_{V,T}$& log$L_{V,img}$
&  log$(M_{\rm bul})$  & $\log (M_{\rm BH})$ 
&log(${M_{\rm BH}\over M_{bul}}$)&$\sigma$(\kms )& $\delta\log L_{bul}$\\
{(1)}&{(2)}&{(3)}&{(4)}&{(5)}&{(6)}&{(7)}&(8)&(9)\\
{}&{}&{}&{}&{}&{}&{}&{}&{}\\
\tableline

{}&{}&{}&{}&{}&{}&{}&{}&{}\\

3C\,120 &$  2.21$& 9.97\tablenotemark{a}  & 10.03  &10.72& 7.49& -3.23 &162&0.52\\
3C\,390.3& 10.5	&10.70&10.18	&10.90& 8.59 & -2.31 & \nodata & 0.52 \\
Akn\,120 &$  5.85$& 10.28& 9.45 \tablenotemark{c} &  $10.04$&8.29&-2.65 & \nodata & \nodata \\
F\,9        &$  5.90$& 10.75& 9.81& 10.46&7.94&-2.52& \nodata & 0.94\\
IC\,4329A   &$ 5.96$& 9.82& 8.86& 9.34  &  $<6.86$ &$<-2.47$& \nodata & 0.97\\
Mrk\,79     &$  6.28$& 9.93&9.54& 10.15& 8.02 &-2.13&125& 0.38\\
Mrk\,110 *  &$  1.67$& 10.16\tablenotemark{a } &9.70\tablenotemark{c}   & 10.74 &  6.91 &-3.83&90&\nodata \\
Mrk\,279&$  3.41$& 10.24 & 9.72\tablenotemark{c}    & 10.36  &  7.41&-2.95& \nodata & \nodata \\
Mrk\,335*   &$  1.26$& 9.86\tablenotemark{a}&  9.89 & 10.56& 6.58 &-3.98& \nodata &-0.03\\
Mrk\,509    &$  2.86$& 10.55& 10.06& 10.76 &	7.98&-2.78& \nodata &0.50\\
Mrk\,590*  &$  2.17$& 10.36& 10.29 & 11.03& 7.15&-3.88&169&0.07 \\
Mrk\,817    &$  4.01$& 10.32& 9.69 & 10.33& 7.56  &-2.77&140& 0.63\\
Mrk\,841        &$  5.41$& \nodata & 9.69 & 10.91& 8.49\tablenotemark{b} &-2.42& \nodata & \nodata\\
NGC\,3227   &$  5.53$& 10.04& 8.96 & 9.47	& 7.69&-1.78&128 & 1.07\\
NGC\,3516   & 2.70&9.61& 10.00\tablenotemark{c}    & 10.69 & 7.36&-3.33 & 124 & \nodata \\
NGC\,3783   &$  4.10$& 9.88&9.15 & 9.69 & 7.04  &-2.65& \nodata & 0.73\\
NGC\,4051*   &$ 1.23$& 9.73& 9.46 & 10.05 & 6.20&-3.85 &80 & 0.27\\
NGC\,4151   &$  5.23$& 9.84&8.83 & 9.31 & 7.08 &-2.23 & 90&1.01\\
NGC\,4253&$4.10	$& \nodata & 9.63 & 10.25& 7.00\tablenotemark{b}  &-3.25& \nodata & \nodata\\
NGC\,4593   & 3.72 & 9.59& 10.20\tablenotemark{c}& 10.21 & 6.91\tablenotemark{b} &-3.30 &124& \nodata \\
NGC\,5548 &$  5.50$& 10.21& 9.68& 10.31 & 7.83  &-2.48&180& 0.53\\
NGC\,6814&$  5.50$& 9.56& 9.25& 9.81 & 7.08\tablenotemark{b}           &-2.73& \nodata & 0.31\\
NGC\,7469   &$  3.20$& 10.21&10.04& 10.74 & 6.88& -3.86& \nodata & 0.17\\
{}&{}&{}&{}&{}&{}&{}&{}&{}\\
\tableline
\end{tabular}
\vspace{-0.3cm}
\tablenotetext{a}{Unknown Hubble type, bulge correction estimated assuming Sa}
\tablenotetext{b} {Bulge magnitude from Virani (1999), BH mass from R-L scaling (eq. 2)}
\tablenotetext{c}{No image, bulge luminosity estimated from FWHM-$\delta L_{\rm bulge}$ 
relation (eq. 1)}
\vspace{-0.7cm}
%
\tablecomments{ 
Narrow line Seyferts are denoted by an asterisk.
Column (2) -- FWHM of (H$\beta$), rms profile, in units of $10^3$\kms. 
(3) --  absolute bulge V luminosity derived from Hubble Type empirical formula from Whittle \et (1992),
(4) -- absolute bulge V luminosity derived by imaging bulge/disk decomposition from MD01,
(5) -- log of the galactic bulge mass ($M_{bul}$) in $\ms$ (MD01), 
(6) -- \bh mass (from WPM), 
(7) -- BH to bulge mass ratio, 
(8) -- stellar velocity dispersion in \kms from Nelson \& Whittle (1995) and Ferrarese \et (2001),
(9) -- log bulge correction ($L_{V,T}/ L_{\rm img}$). 
}
\end{center}
\end{table*}
\begin{table*}
\begin{center}
\vspace{-0.5cm}
\scriptsize
\caption{Quasar data\label{tbl-2}}
\begin{tabular}{llllll}
\tableline\tableline
Name& FWHM &   log$L_{V}$&  log$(M_{\rm bul})$ & $\log (M_{\rm BH})$ 
&log(${M_{\rm BH}\over M_{bul}}$) \\
{(1)}&{(2)}&{(3)}&{(4)}&{(5)}&{(6)} \\
\tableline
{}&{}&{}&{}&{}&{}\\
0052+25 &	4.37	 & 10.41	 & 11.18	 & 8.29	 & -2.89  \\
0054+14  &	9.66	 & 10.65	 & 11.46	 & 8.91	 & -2.55  \\
0137+012 & 	7.61	 & 10.81	 & 11.65	 & 8.57	 & -3.08  \\
0157+001*&	2.14	 & 10.93	 & 11.79	& 7.70	 & -4.09 \\
0204+292 &	6.80	 & 10.53	 & 11.32	 & 7.15	 & -4.17 \\
0205+02 &	 2.90	 & 9.65	 & 10.28	& 7.86	 & -2.42 \\
0244+194 &	3.70	 & 10.33	 & 11.08	 & 8.03	 & -3.05 \\
0736+017 &	2.96	 & 10.65	 & 11.46	& 7.99	 & -3.47 \\
0923+20 &	7.30	 & 10.53	 & 11.32	 & 8.94	 & -2.38 \\
0953+41 &	2.96	 & 10.33	 & 11.08	 & 8.39	 & -2.69 \\
1004+13 &	6.34	 & 10.85	 & 11.70	& 9.09	 & -2.61 \\
1012+008 &	2.64	 & 10.77	 & 11.60	& 7.79	 & -3.81 \\
1020-103 &	7.95	 & 10.57	 & 11.36	 & 8.35	 & -3.01 \\
1029-14 &	7.50	 & 10.41	 & 11.18	& 9.07	 & -2.11 \\
1116+215 &	2.92	 & 10.69	 & 11.51	 & 8.22	 & -3.29 \\
1202+28 &	5.01	 & 10.33	 & 11.08	 & 8.29	 & -2.79 \\
1217+023 &	3.83	 & 10.73	 & 11.55	 & 8.40	 & -3.15 \\
1226+02 &	3.52	 & 10.97	 & 11.84	 & 8.61	 & -3.23 \\
1302-10 &	3.40	 & 10.61	 & 11.41	 & 8.30	 & -3.11 \\
1307+08 &	5.32	 & 10.37	 & 11.13	& 8.05	 & -3.08 \\
1309+35 &	2.94	 & 10.41	 & 11.18	& 7.99	 & -3.19 \\
1402+26*&	1.91	 & 9.93	 & 10.61	& 7.28	 & -3.33 \\
1425+267\tablenotemark{a} &	9.40	 & 10.57	 & 11.36	& 9.32	 & -2.04 \\
1444+40 &	2.48	 & 10.37	 & 11.13	 & 8.06	 & -3.07 \\
1545+21 &	7.03	 & 10.53	 & 11.32	 & 8.93	 & -2.39 \\
1635+119 &	5.10	 & 10.49	 & 11.27	 & 8.10	 & -3.17 \\
1704+60*\tablenotemark{b} &	0.40& 10.37	 & 11.13	& 6.87	 & -4.86 \\	
2135-14 &	5.50	 & 10.61	 & 11.41	 & 8.94	 & -2.47 \\
2141+175& 	4.45	 & 10.73	 & 11.55	 & 8.74	 & -2.81 \\
2247+140*&	2.22	 & 10.73	 & 11.55	 & 7.59	 & -3.96 \\
2349-01 &	5.50	 & 10.97	 & 11.84	 & 8.78	 & -3.06 \\
2355-082 &	7.51	 & 10.65	 & 11.46	 & 8.39	 & -3.07 \\
{}&{}&{}&{}&{}&{}\\
\tableline
\end{tabular}
\vspace{-0.3cm}
\tablenotetext{a}{Kirhakos \et 1999}
\tablenotetext{b}{FWHM(rms) and $M$(rms) from Kaspi \et (2000)}
\vspace{-0.5cm}
\tablecomments{Narrow-line quasars are denoted by an asterisk.
Column (2) -- FWHM of (H$\beta$) in units of $10^3$\kms . 
(3) --  absolute bulge V luminosity (in units of $L_\odot$,from MD01),
(4) -- log of the galactic bulge mass,
(5) -- \bh mass, 
(6) -- BH to bulge mass ratio
}
\end{center}
\end{table*}

The virial formula for the BH mass in AGNs requires the size and gas velocity of the broad
emission line region. 
Most of the quasars do not have \rev data. For those we estimate the size of the BLR using
 the empirical correlation 
$$R_{BLR}=33 (\nu L_\nu (5100 )/10^{44} \ers )^{0.7} {\rm light-days}\eqno (2)$$
between the BLR size and the continuum luminosity (Kaspi \et 2000).
We test the goodness of the empirical R-L relation for the quasars with \rev data.
There are four quasars in the MD01 sample with \rev data (Kaspi \et 2000) - 
PG0052, PG0953, 3C273 and PG1307.
Comparing the BH masses
the agreement between the empirical R-L relation and the \rev result
is within a factor of 1.5.
 
There are three quasars for which our BH mass (calculated with the empirical formula)
differs significantly from the values given by MD01 (after correcting for their
different kinematic factor, see 2.4 below): 
PG1307 (the mass in MD01 is by a factor 4 too low, apparently because a mistaken FWHM), similarly for
0204+292 (3C059) MD01 get a BH mass too low by a factor 40 (they have FWHM=1040 km/s, while we use a FWHM(H$\beta$) of 6800 km/s, cf. Erackleous \& Halpern 1994) 
and PG 1012+008 (their BH mass is by a factor 10 too low, probably a typo).

\subsection{Quiescent Galaxies}
For comparison with inactive or weakly active galaxies we adopt the sample of Kormendy 
\& Gebhardt (2001), which lists 37 galaxies with BH mass determinations from stellar or gas 
dynamics and bulge magnitudes. We use the relation B-V=0.8 in order to translate
the B bulge magnitude to our standard V luminosity. We exclude the two galaxies NGC4486B and NGC5845 
which have a large uncertainty (factor 10) in the BH mass, significantly larger
than the other objects in the sample. In the $M_{\rm BH}-\sigma$ analyses we also exclude NGC 4945
which does not have a $\sigma$ observation.

\subsection{Black Hole Mass}
The BH mass of AGNs can be estimated assuming the emission-line gas is bound (see section 6.5). 
The virial relation the gives $M=k^2G^{-1}R_{blr}v_{FWHM}^2$, where $kv_{FWHM}$ is the velocity dispersion
in the gas (deduced from the observed width $v_{FWHM}$ of the emission lines)
and the kinematic factor $k$, which depends on the kinematics, geometry and emissivity of the BLR.
Assuming the velocity dispersion in the line emitting gas is isotropic we have
$k={\sqrt 3 \over 2}$,
which gives
$$M_{\rm BH}=1.46 \ms \times 10^5 \left ({R_{blr}\over {\rm lt-days}}\right ) 
\left ({v_{FWHM}\over 10^3\kms }\right )^2\eqno (3)$$
Non isotropic geometry and velocity dispersion (combined with a non-uniform  distribution of inclinations)
can give different values of the coefficient $k$ (see section 6.4).
MD01 assume a flattened BLR geometry and a distribution of inclination biased to face-on,
which yield $k=3/2$ and hence the virial BH mass is 3 times larger than in the isotropic geometry.

\section{The BH-bulge Relation}

\subsection{The Luminosity Relation}

In fig. 2 we plot the BH mass vs. bulge luminosity.
Notably the scatter in the Seyfert and quasar groups is quite large. 
We note however that all AGNs with low $M_{\rm BH}/L_{\rm bulge}$ values
are narrow \el objects (NLS1s or narrow line quasars).  
In other words, it seems that narrow line AGNs tend to have small BHs compared 
to their host bulges. 
We draw the limit between narrow and broad line Seyferts at 2200\kms,  
which is close to the traditional NLS1 definition (2000\kms ) but includes a couple
of narrow line objects in our sample which are very close to that definition 
(MRK 590, PG0157+00 and PG2247+14).
The precise value is not important, as the NLS1 definition is merely a historic convention,
and our results are not very sensitive to the exact border value. 

In section 4
we show  that there is a strong correlation between line width and BH/bulge ratio,
so that the transition from broad to narrow line objects is gradual, as well as the transition
from high to low BH/bulge ratios.
This correlation further supports the separation of the narrow-line objects:
as the BH mass estimate of AGNs strongly depends on the line width, removing the narrow-line 
objects may seem to automatically remove the outlying data points with low BH/bulge ratio.
However, we find that this is not the case, as the BH/bulge mass ratio depends also on  
other observables: the BLR radius and the bulge magnitude.
In order to test the significance of separating the narrow line objects, we compare the two correlations:
BH mass vs. FWHM and BH/bulge ratio vs. FWHM. The correlation coefficients are
0.62 and 0.80, respectively, which shows that the BH/bulge - FWHM relation is more significant than the mere BH mass dependence on the FWHM.
 
In section 6 we discuss a few physical and observational models which could explain 
such a correlation and motivate the separation of the narrow line objects.
We group those objects (Narrow Line Seyfert 1s and narrow line quasars) as a 
separate class which we denote NLAN (Narrow Line Active Nuclei). 

\begin{inlinefigure}
\centerline{\includegraphics[width=1.2\linewidth]{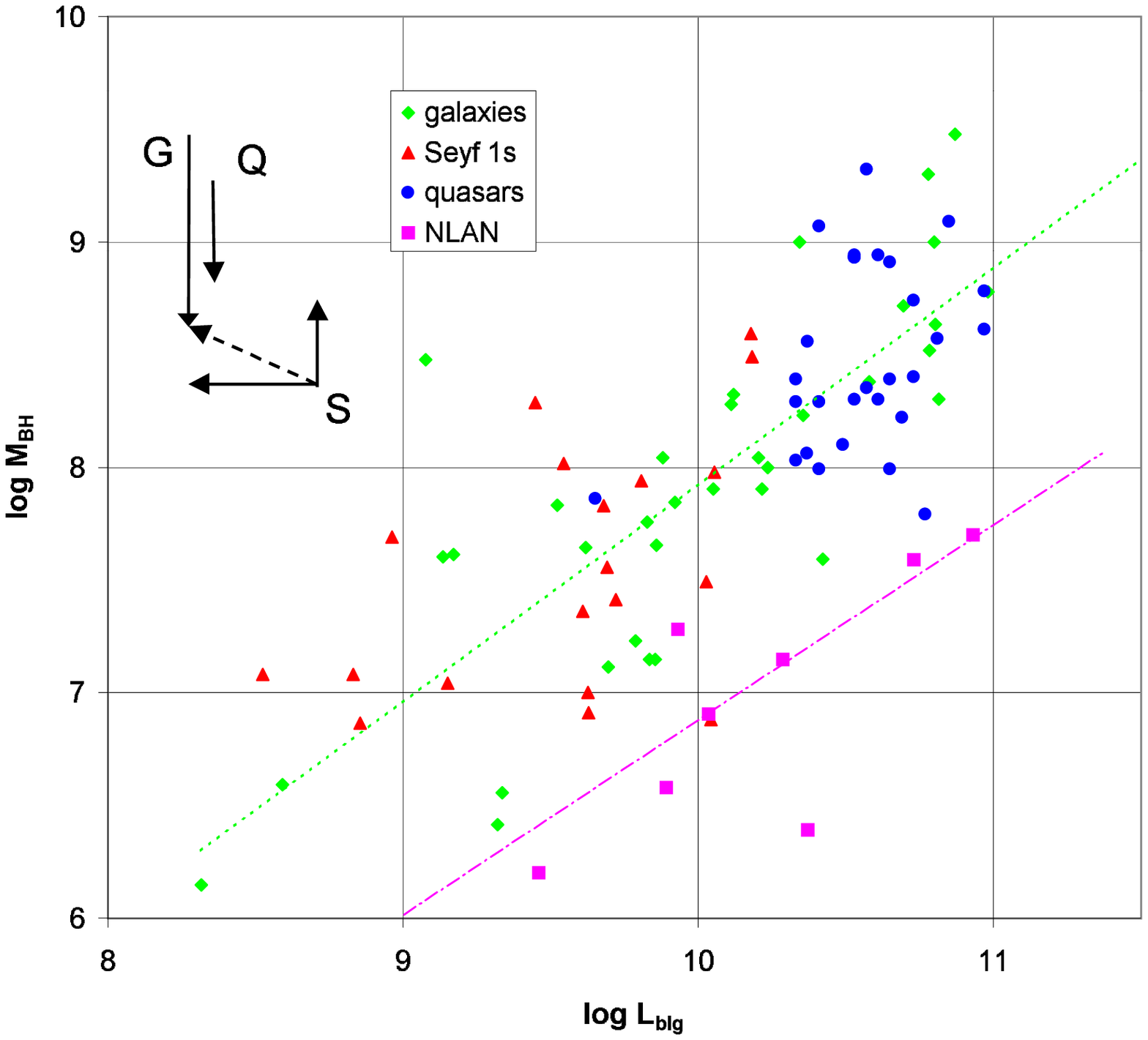}}
\figcaption{
The BH mass vs. bulge luminosity. Green diamonds denote BHs detected by stellar and gas 
dynamics in quiescent galaxies (Kormendy  \& Gebhardt 2001), 
blue 
circles - quasars (MD01), red 
triangles - Seyfert 1 galaxies (BH mass from WPM,
bulge luminosity from MD01) and pink 
squares - Narrow Line Active Nuclei (NLS1s from WPM and NLQs from MD01).
The least-squares linear fit for quiescent galaxies is indicated by
a dashed green 
line and NLANs by a pink, 
dot dashed line.
The arrows indicate the effects of the improvements in the data that cancelled the apparent
former discrepancy in the $M_{BH}/L_{\rm bulge}$ ratio between Seyferts and quiescent galaxies and quasars (see text).
}
\end{inlinefigure}

The best fits to the data are ($L$ and $M$ are in solar units)
\begin{enumerate}
\item
Quiescent galaxies (35):

$\log M= (0.96\pm 0.13) \log L -(1.7\pm 1.3)$, 
\item
Broad line AGNs (19 Seyferts and 28 quasars) 

$\log M = (0.90\pm 0.11) \log L -(1.1\pm 1.1)$,

\item
Narrow line AGNs (4 Seyferts and 4 quasars):

$\log M = (0.84\pm0.21) \log L -(1.5\pm 2.2) $
\item
Quiescent galaxies and broad line AGNs (82)

$\log M = (0.93\pm0.13) \log L -(2.9\pm 2.6) $
.
\end{enumerate}

The fits for the BH/bulge mass relation, along with the corresponding correlation
measures are detailed in table 3 below.  
The (broad line) AGNs have an almost linear fit
very similar to that of inactive galaxies and to
the combined sample of broad-line AGNs and quiescent galaxies. 
The NLAN group has a similar slope but with an $M/L$ ratio 
approximately a factor of 10 lower.

\subsection{The Mass Relation}

Transforming bulge luminosity to mass one finds a similar correlation between the BH mass and
the bulge mass. The correlation is analogous to the
BH mass vs. bulge luminosity relation (fig. 2), with a somewhat less steep slope (since $M_{\rm bulge}\sim L^{1.2})$.
Fig. 3 shows the BH mass vs. the bulge mass. 
In we list the best fits to the $M_{\rm BH}$ vs. $M_{\rm bulge}$ relation for
broad line AGNs (Seyfert 1s and quasars), 
narrow line AGNs (NLS1s and NL QSOs) and quiescent galaxies, as well as for the combined sample.
For each subset we give the three measures of the fit:
the correlation coefficient, the standard deviation of the (logarithmic) scatter, given by 
$dM=[\Sigma(\delta\log (M_{bh})^2/(N-1)]^{1/2}$
and the reduced $\chi^2$ of the fit. 
Since the BH mass estimates have systematic modeling uncertainties (section 4), 
we approximate the uncertainty in the $\chi^2$ calculation by a uniform error of 0.5 dex 
(factor 3) in the BH mass estimate, which is derived by combining the typical observational 
error with the systematic uncertainty (see section 3.3 below).

We also list the mean BH/bulge mass ratios and their standard deviations.
In order to compare the BH-bulge correlation with that of the $M_{\rm BH}-\sigma$ relation we
have added a row showing the $M_{\rm BH}-\sigma$ correlation for quiescent galaxies. 
Since both correlations have the same y-variable ($M_{\rm BH}$) they can be compared using the same statitstic.
For consistency we treat the errors in the BH mass for AGNs and quiescent galaxies in the same manner, although
in some cases, such as the Galaxy and NGC 4258, the error in the BH mass is much smaller.
Note that the effective error (factor 3) assumed for the BH mass is larger than the measurement uncertainty for most BHs in quiescent galaxies. Using merely the measurement uncertainty in $M_{\rm BH}$, e.g. the error bars quoted in the Kormendy and Gebhardt (2001) sample gives $\chi^2=$3.5
for the $M_{\rm BH}-\sigma$ relation and $\chi^2=10.5$ for $M_{\rm BH}$-bulge relation.

We note several important results:
\begin{enumerate}
\item
Broad-line (BL) AGNs have a very similar distribution (for both, slope and mean MH/bulge ratio) to that
of the quiescent galaxies.  
\item
In spite of their small number in out sample, removing the narrow line AGNs has a significant effect
on the combined AGN sample: the total AGN sample has a significantly lower correlation coefficient,  
and a lower mean BH/bulge ratio than BL AGNs and quiescent galaxies.  
\item 
NLANs have an average a BH/bulge ratio lower by a factor of 10
than the broad line AGN average, and the normalization of the best fit is similarly lower.
\item
Quasars and Seyfert galaxies as separate groups have shallower slopes and lower 
correlation coefficients, due to the lower dynamical range
of each of these groups separately.
\item
The BH-bulge correlation of broad line AGNs is quite tight, not much less than
the $M_{\rm BH}-\sigma$ relation for quiescent galaxies.
\end{enumerate}

The last item is demonstrated by comparing the three correlation statistics in table 3
(the correlation coefficient, the logarithmic scatter and $\chi^2$) for the BH-bulge and the BH-$\sigma$ relations. 
For the 47 BL AGNs we find 0.76, 0.43 and 0.78, respectively, nearly as tight as the $M_{\rm BH}-\sigma$ correlation of the 34 quiescent galaxies in the Kormendy and Gebhardt (2001) sample (0.86, 0.38 and 0.60). 
We note that excluding the Seyfert galaxy
NGC 7459 (which has an exceptionally low BH mass relative to its luminosity) the BH-bulge correlation of the broad-line AGNs becomes 
even tighter (dM=0.39).

\begin{inlinefigure}
\centerline{\includegraphics[width=1.15\linewidth]{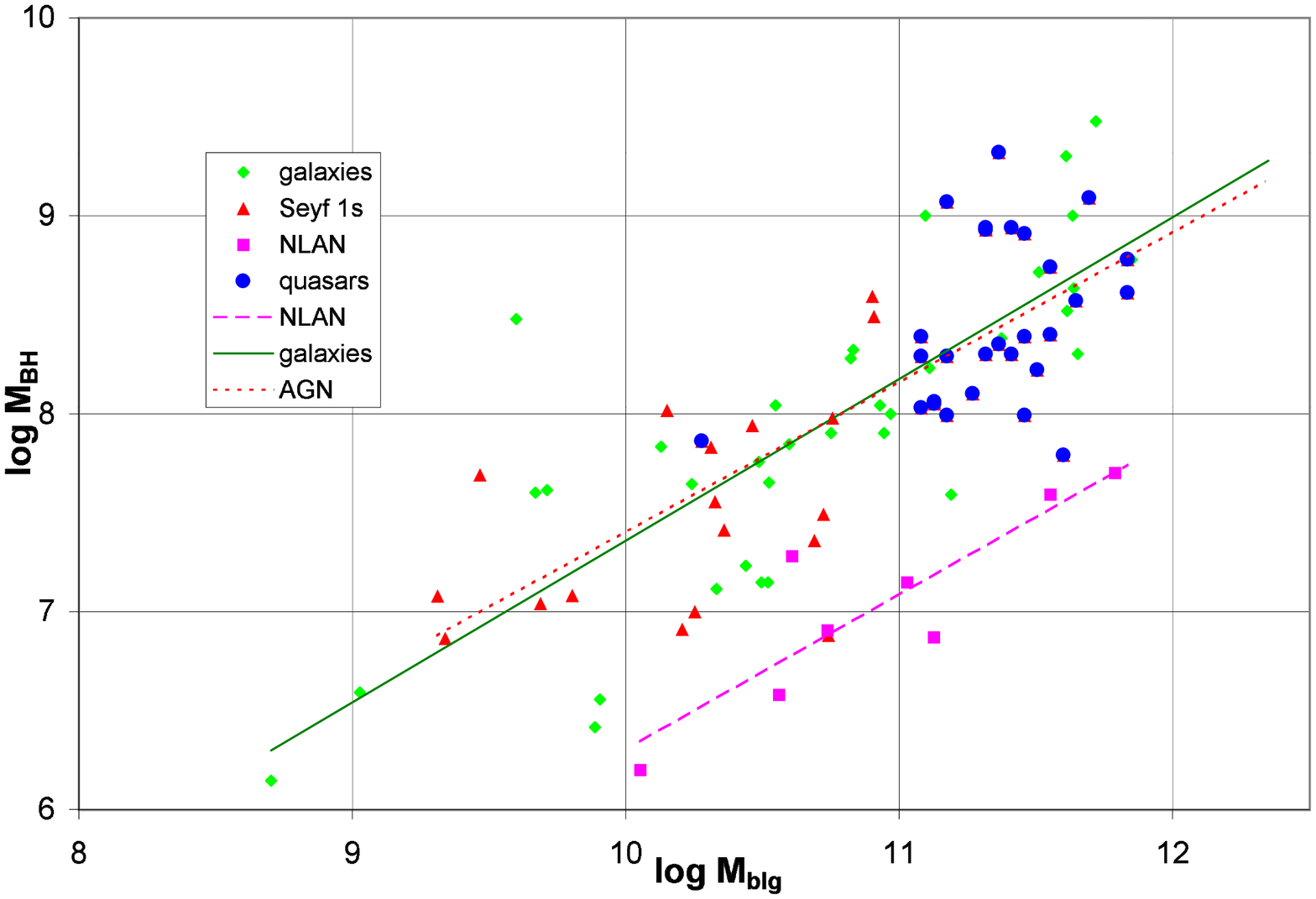}}
\figcaption
{Mass estimates of MBHs plotted against the
mass of the host bulge (or host galaxy for elliptical galaxies). Green 
Diamonds denote BHs in inactive galaxies, blue 
circles - quasars, red 
triangles - Seyfert 1 galaxies and pink 
squares - Narrow Line Active Nuclei (NLS1s and NLQs).
The least-squares linear fits for each class are indicated by
a solid green 
line (quiescent galaxies), dashed red 
line (broad line AGNs)
and a pink, 
long dashed line (NLANs).}
\end{inlinefigure}

\begin{table*}

\begin{center}
\scriptsize
\caption{BH-bulge Mass Correlation's\label{tbl-3}}
\begin{tabular}{lccccccc}
\tableline\tableline
{}&{}&{}&{}&{}&{}&{}&{}\\
Set& N & a & b & $M_{\rm BH}/M_{\rm blg}$ & R & $\delta M$ & $\chi^2$ \\
{}&{(1)}&{(2)}&{(3)}&{(4)}&{(5)}&{(6)}&{(7)} \\
\tableline
{}&{}&{}&{}&{}&{}&{}&{}\\
All AGNs & 55 & 0.74$\pm$0.11 &  -0.09$\pm$1.22 & -2.93$\pm$0.57 & 0.67 & 0.54 &1.20 \\ 
{}&{}&{}&{}&{}&{}&{}&{}\\
Broad Line AGNs & 47 & 0.73$\pm$0.09 &  -0.15$\pm$1.02 & -2.80$\pm$0.48 & 0.76 & 0.43 &0.78 \\ 
{}&{}&{}&{}&{}&{}&{}&{}\\
Quiescent Galaxies& 35& 0.82$\pm$0.11& -0.82$\pm$1.15& -2.77$\pm$0.50 & 0.80& 0.48&0.96 \\
{}&{}&{}&{}&{}&{}&{}&{}\\
BL AGNs and Q.Galaxies & 82 & 0.79$\pm$0.07 &  -0.53$\pm$0.74 & -2.79$\pm$0.47 & 0.79 & 0.45 &0.82 \\ 
{}&{}&{}&{}&{}&{}&{}&{}\\
\tableline
{}&{}&{}&{}&{}&{}&{}&{}\\
BL quasars & 28 & 0.51$\pm$0.24 & 2.70$\pm$2.74& -2.86$\pm$0.41 & 0.38 & 0.38& 0.61 \\
{}&{}&{}&{}&{}&{}&{}&{}\\
BL Seyfert 1s &19 & 0.52$\pm$0.23 &  2.27$\pm$2.37 & -2.68$\pm$0.55 & 0.48 & 0.49 &1.01 \\ 
{}&{}&{}&{}&{}&{}&{}&{}\\
NLANs & 8 & 0.78$\pm$0.17 & -1.53$\pm$1.9& -3.9$\pm$0.27 & 0.88 & 0.24 & 0.27 \\
{}&{}&{}&{}&{}&{}&{}&{}\\
\tableline
{}&{}&{}&{}&{}&{}&{}&{}\\
Quiescent Galaxies-$\sigma$ & 34 & 3.65$\pm$0.37 &  -0.29$\pm$0.85 & 5.71$\pm$0.60 & 0.86 & 0.38 &0.60 \\ 
{}&{}&{}&{}&{}&{}&{}&{}\\
\tableline
\end{tabular}

\tablecomments{Columns:
(1) N  - number of objects, (2-3) coefficients of the linear fit 
$\log(M_{\rm BH})=a\log (M_{\rm bulge})+b$
and standard deviations,
(4) mean log(BH/bulge) mass ratio and standard deviation, (5) R - correlation coefficient, 
(6) Standard deviation scatter of the fit $\delta(\log (M_{\rm BH}))$ 
(7) - reduced $\chi^2$ assuming the error in $\log (M_{\rm BH})$ is 0.5 dex.
 }
\end{center}
\end{table*}

\subsection {Uncertainties}
The formal uncertainty in the slope of the linear fit to the
(broad line) AGNs is 0.09. 
However, for a realistic uncertainty it is necessary 
to take into account the 
errors in the BH mass estimate and bulge luminosity.
The errors in the BH mass are of two kinds:
systematic errors in the \rev mapping 
and virial mass calculation and measurement
errors. The latter are random and typically 
of the order of 2-3 (see WPM for the observational
uncertainty in the \rev masses and for 
the error of the $L-R$ scaling method 
(used in our sample for most of the
quasars) compared to the \rev technique ).
The systematic errors are not random, but 
depend on the geometry and dynamics of the broad emission 
line region, and in extreme cases could 
amount to a factor of 3 in each direction (e.g. Krolik 2001; 
see section 6 below). However, since this 
dependence is largely unknown and can act in both directions, 
depending on the conditions in each AGN, 
we treat the systematic errors as random ones as well.
 
The uncertainty of a linear fit 
$y=ax+b$ for N data pairs $(x,y)$ with typical errors $\delta x$ and $\delta y$ may be estimated as
$\delta a = [  \delta a_{\rm formal}^2 + (a \delta x /\sqrt N)^2 + (\delta y /\sqrt N )^2 ]^{1/2}$.
Combining the formal error in the fit with the random 
measurement errors in the BH mass (assumed to be a factor of 3) and bulge  
luminosity (factor 2), and the systematic errors in the BH mass estimate
(assumed to be random with a factor of 3)
for the broad-line AGN sample (N=47), gives $\delta a =0.16$, so that 
$$M_{\rm BH}\propto L_{\rm bulge} ^{0.90\pm 0.16}\eqno (4a)$$
and
$$M_{\rm BH}\propto M_{\rm bulge} ^{0.76\pm 0.15}\eqno (4b)$$

with  $<M_{\rm BH}/ M_{\rm bulge}>=0.0015$.

Studying a sample of 16 AGNs Laor (2001) finds a steeper fit,  
$M_{\rm BH}\propto M_{\rm bulge} ^{1.53\pm 0.14}$ (formal error only),
which is $\sim 5-6\sigma$ higher than what we find in this work
($M_{\rm BH}\propto M_{\rm bulge} ^{0.73\pm 0.09}$).
This difference is explained as follows.
(i) Laor's fit includs quiescent galaxies, 
some of which have a low bulge luminosity
and a low $M_{\rm BH}/M_{\rm bulge}$ value (see section 6.2). 
For the AGNs separately Laor finds a less steep slope of 1.36$\pm$0.15. 
Furthermore, a single point significantly influences the slope -
the  NLS1 NGC 4051 - which extends the
dynamical range by almost an order of magnitude. If it were excluded
(being a narrow line object), the slope of the fit would decrease
considerably (and the standard deviation in the slope would increase).
Excluding NGC 4051 and adding the measurement error would flatten the slope 
and increase the uncertainty in Laor's fit, which would ecome consistent
with linearity. 

 (ii) The BH masses used by Laor (1998; 2001) 
are systematically larger than those used in this work (and in MD01).
The BH masses in Laor's sample are are on average larger by a factor of 1.5
than those calculated for the same quasars in our sample. 
This discrepancy originates in part from 
the different luminosity measures used: Laor used the 
broad band (0.1-1 micron) luminosity (Neugebauer et. 1979) while we 
(and MD01) use the monochromatic luminosity $\nu F_\nu (5100)$. 
The latter seems more 
adequate here, as we (as well as MD01) 
estimate the BH mass of the quasars using the
empirical $R-L$ relation of Kaspi \et (2000)
which has been derived with  $\nu F_\nu (5100)$. 


We conclude that quasars, Seyfert galaxies and inactive galaxies 
have the same BH-bulge relation and BH/bulge ratio, while NLANs 
seem to have BH/bulge ratios lower by factors of $\sim$1-30.

\subsection{Resolving the discrepancy: Do Seyferts have a lower BH/bulge ratio than Quiescent Galaxies?}

Previous works have found a significant difference in 
the \bh mass- bulge luminosity (or mass) relationship of Seyferts and ordinary galaxies, the former showing systematically lower $M_{\rm BH}/L_{\rm bulge}$
values (Wandel 1999, Ho 1999, Gebhardt 2000b). 
The discrepancy between the $M_{\rm BH}/M_{\rm bulge}$ 
ratio of quiescent galaxies and Seyferts (factor of 10-20 on average) was a result of three factors, shown schematically by the arrows in fig. 2: 
(i) the Magorrian BH masses were overestimated by an average 
factor of $\sim 5$ ($M_{\rm BH}/M_{\rm bulge}=$0.6\% 
compared with the current estimate of 0.1-0.13\% 
(Ho 1999; Merritt \& Ferrarese 2001a). The average decrease in the quiescent galaxy BH mass
is shown in fig. 2 by the arrow denoted G. 
(ii) the Seyfert bulge luminosity calculated using the Simien de Vaucouleurs formula is 
too large (by a factor of $\sim 2-3$); the
bulge luminosity estimated by direct imaging and bulge/disk decomposition is lower by an average 
factor of $\sim$2 (0.15dex in our total AGN sample, as seen in table 3, 
but more when applied to smaller sub-samples, e.g. to the Seyferts) represented by the horizontal arrows.  
(iii)
Narrow Line Seyfert 1 galaxies, which do seem to have BH/bulge 
mass ratios lower than  broad line active galaxies, lowered the average.
Excluding the NLS1s raises the average $M_{\rm BH}/M_{\rm bulge}$ value of AGNs by a factor of
$\sim 2$, denoted in fig. 2 by the arrow pointing upwards.
Items (ii-iii) combined increased the BH/bulge ratio of Seyferts by an average factor of $\sim$3.5.
(dashed arrow).   
Similarly, the disagreement between the BH/bulge ratios of the PG quasars and \rev Seyferts (Wandel 1999)
disappeared due to the revision of the Seyferts' BH/bulge ratio 
and the lower BH 
mass estimates of the quasars (the arrow denoted by Q in fig 2).

Given the present good agreement of AGNs and quiescent galaxies, one may ask whether there is room for
systematically higher BH mass estimates in AGNs, e.g. due to inclination effects. In the geometry suggested by MD01, the virial
BH mass of AGN would be larger on average factor of 3, compared with the mass calculated under the assumption
of isotropic velocity distribution. We can check what would be the implication of such an increase on the
BH-bulge and BH-velocity dispersion relations of AGNs compared with those of quiescent galaxies (where the BH mass estimates are not affected). 
In order to test this hypothesis
we calculate the $\chi^2$ of the AGN sample 
(including the NLANs) with the BH mass increased by a 
uniform factor 3, with respect to  
the best fit to the quiescent galaxies presented in table 3.
Assuming the typical measurement + modeling error in the BH mass is 0.5 dex, this gives a reduced $\chi^2$ of 2.2
compared with 1.20 for BH masses calculated with an "isotropic BLR" 
assumption for all AGNs in our sample. 
A similar exercise can be done for the $\sigma-M_{BH}$
relation, which has less AGNs, but also lower spread 
and a smaller error in the BH mass. 
Increasing each BH mass by a factor of 3, 
we find for the 11 Seyfert galaxies with velocity  
dispersion data (table 1)  $\chi^2$ of 2.58 compared with 1.08 for the "isotropic BLR"  BH masses. We conclude that the BH mass calculated under the "isotropic BLR" assumption for the BLR geometry is more consistent with the BH-bulge and BH-velocity dispersion relations for quiescent galaxies. 
 
\section {The BH-bulge vs. \el width correlation}

We find that narrow-line AGNs - 
NLS1s and narrow line quasars - tend to have low $M_{\rm BH}/M_{\rm bulge}$ values. 
This gives a physical motivation to the exclusion of the narrow line AGNs as a separate group in the
BH-bulge relation.

Mathur \et (2001) make a similar suggestion based on a sample of 15 NLS1s, 
but with less reliable  BH masses. 
Only three of the objects in their sample have
\bh masses measured by \rev , 3 more by the scaling-virial method, and 
the 9 remaining  BH masses are calculated using accretion disk modeling, which is less reliable.
They find that their accretion disk model gives BH masses too large 
(by a factor of 3-7)
compared with the \rev masses (where available). In order to compensate for this discrepancy
they decrease the  BH masses given by the accretion-disk model by a calibration 
correction factor of 5.
For the bulge magnitude Mathur \et
use  Whittle (1992) with the Simien-deVaucouleurs relation 
(which is likely to overestimate the bulge magnitude in AGN, as discussed in above).  

Considering all $55$ AGNs in our sample with available $M_{\rm BH}$ and bulge magnitude 
we find that the smaller 
BH/bulge ratio of narrow line objects is in fact the lower end of a more general relation:
the BH mass /bulge luminosity ratio is strongly correlated with the \el width.

\begin{inlinefigure}
\centerline{\includegraphics[width=1.2\linewidth]{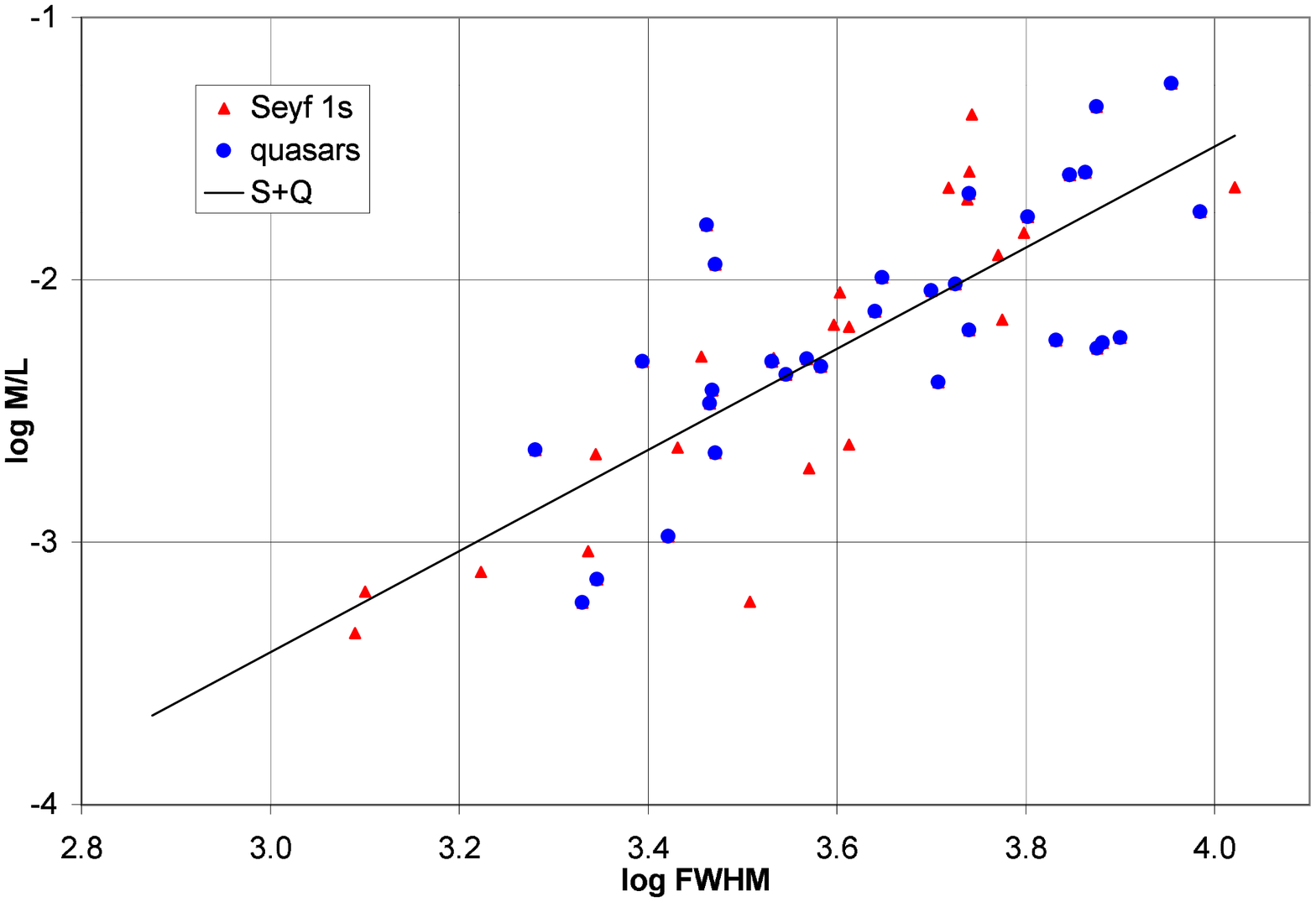}}
\figcaption{BH-to-bulge luminosity ratio  plotted against the
H$\beta$ broad \el width for AGNs.Red 
Triangles represent Seyfert 1 galaxies and blue 
circles are quasars.
The solid line indicates the least-squares linear fit.}
\end{inlinefigure}

In fig. 4 we plot $M_{\rm BH}/L_{\rm bulge}$ vs. FWHM(H$\beta$).
The best linear least-squares fit is

$$\log \left ({M_{\rm BH}\over L_{\rm bulge}}\right )_\odot =
1.94 \log \left ( {{\rm FWHM(H}\beta)\over
10^3\kms}\right ) - 3.40 ,\eqno (5)$$
with a correlation coefficient $R=0.80$.
BH mass and bulge luminosity are in solar units, and we have excluded PG1704 because
of its highly uncertain reverberation BH mass determination (however, if we take the 
BH mass and FWHM values quoted in Kaspi \et (2000) it agrees well with the fit in fig. 4).

The tight correlation (5) is not surprising - it is actually 
an artifact of the virial relation: since $M_{\rm BH}\propto v(FWHM)^2$,
this  correlation merely means that $R_{BLR}/L_{\rm bulge}$ is independent on the line width.
Combining eqs. (3) and (5) we have
$$R_{BLR}\approx 27 \left ( {L_{\rm bulge}\over 10^{10} L_\odot}\right )\quad {\rm lt-days}.$$ 
Looking at the data we find almost precisely the same result  - 
a very tight correlation between bulge luminosity and BLR size
(in preparation), as shown in fig. 5. 

The best fit to all 55 AGNs in our sample combined is
$\log (R/{\rm lt-days})=1.05 \log (L_{\rm bulge}/L_\odot ) -8.92 $ or
$$R= 13.5 L_{10}^{1.05}\quad {\rm lt-days} \eqno (6)$$
with a correlation coefficient of 0.91.
We note that while Seyferts and NLS1s show a significant correlation, quasars alone do not. 
This is not surprising,
as the BLR size of most quasars in our sample is determined from the R-L scaling relation,
which is less reliable than \rev mapping, used 
for most of the Seyferts. 
However, the quasars, which do have \rev measured BLR 
sizes (solid circles in fig. 5) are indeed very close to the fit.

\begin{inlinefigure}
\centerline{\includegraphics[width=1.15\linewidth]{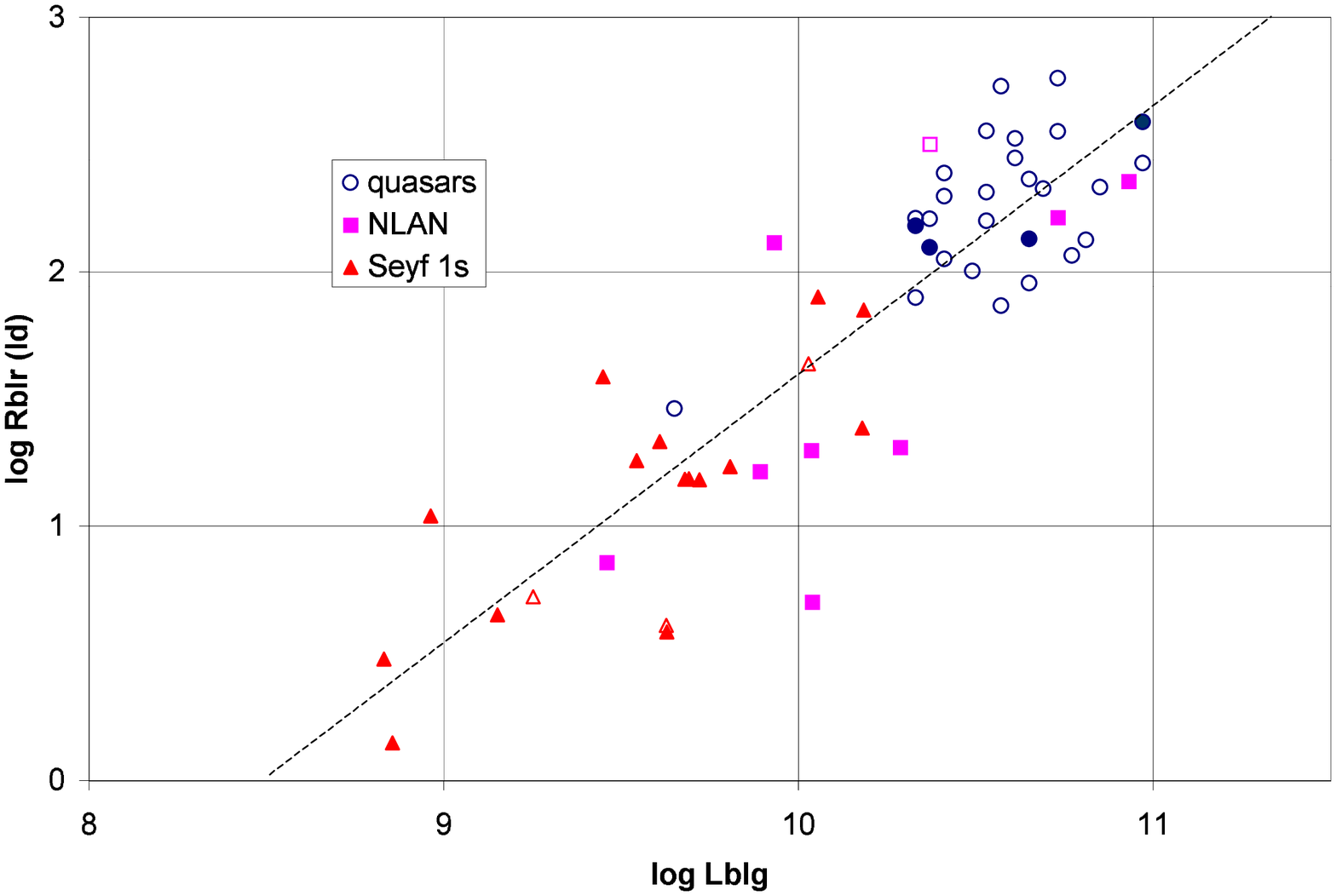}}
\figcaption
{The size of the broad emission-line region vs. the bulge luminosity.
Red 
Triangles are Seyfert 1 galaxies, blue 
circles are quasars and pink 
squares are narrow line AGNs.
Solid symbols indicate BH masses calculated from \rev-mapping data (from WPM and Kaspi \et 2000) and 
open symbols - calculated from the L-R empirical relation.
The solid line indicates the least-squares linear fit.}
\end{inlinefigure}

This is a non-trivial
correlation, not just a derivation,
as it relates two independent observables: on the one hand
the bulge luminosity, a global galactic property on a kpc-scale,
and on the other hand
the BLR size (measured by \rev mapping or luminosity scaling), on
a few light-days scale.

It may be reflecting the BH-bulge relation, but the correlation being
much stronger than the BH-bulge relation supports the case that this
new relation is more fundamental.
A possible connection to basic physical properties 
may be implied by combining the new BLR-bulge relation, the
empirical relation between the BLR size and the central source
luminosity (e.g. Kaspi \et 2000) and the BH-bulge relation
giving that the Eddington ratio increases with BH mass.  

\section{The \bh mass - bulge velocity dispersion relation for AGN}
We have shown that the \bh /bulge ratio in NLS1 
galaxies seems to be significantly smaller 
than in broad line AGNs and in quiescent galaxies.
However, Seyfert galaxies seem to be consistent 
with the $\sigma -M_{\rm BH}$ relation of inactive 
galaxies (Gebhardt \et 2000b; 
Merritt \& Ferrarese 2001b; Ferrarese \et 2001).
In order to unveil the origin of the lower BH/bulge ratio 
of NLANs (namely, whether they really
have lower BH masses 
or rather larger bulges, 
relative to the broad line AGN) 
we look at their 
location in the $\sigma -M_{\rm BH}$ plane.
The bulge velocity dispersion is measured only 
for 3 NLS1s (Mrk 110, Mrk 590 and NGC 4051) and 8
broad-line Seyferts (table 1).

We may use the 
Faber-Jackson relation to estimate the velocity 
dispersion of other AGNs with measured or estimated 
bulge luminosity. Also the relation between the 
the narrow emission lines and the velocity
dispersion (Nelson 2000) may be used to
estimate the velocity dispersion in the bulge.
Here we use the standard F-J relation, e.g.
$L= L_o  \sigma_2^4$,
where $\sigma_2=\sigma/100\kms$ and $L_o$ is a luminosity coefficient determined
by a linear fit with a slope of 4 to the observed Seyfert
Galaxies (fig. 6).

In our limited data set we note that NLS1s have a significantly larger luminosity coefficient 
than broad line ones:
while for broad line Seyfert 1 galaxies $L_o=10^9\ers$, for NLS1s galaxies $L_o=6\times10^9\ers$. 
In addition, galaxies with a massive BHs and AGNs in particular, may have a 
different (flatter) F-J relation than galaxies in general (Wandel 2001 and work in preparation).

Using these relations one may compile a larger sample and increase the dynamical range by including 
also quasars, to estimate the $\sigma -M_{\rm BH}$ relation for AGNs.
In fig. 7 the BH mass is plotted against the stellar velocity dispersion. We see that AGNs are 
consistent with the $M_{\rm BH}-\sigma$ relation of inactive galaxies 
($M_{\rm BH}\propto\sigma^\alpha$, with  
$\alpha=3.5-5$; Gebhardt 
\et (2000a) find $\alpha=$3.65
and  Merritt and Ferrarese (2001a) give 4.72). Adding the quasars seem to favor a steeper slope, but 
this result merely reflects the assumed $L\propto \sigma^4$ relation; if, for example, a flatter 
$\sigma -L$ relation is 
assumed, the luminous objects would have a larger estimated velocity dispersion yielding a flatter slope 
for the $\sigma-M_{\rm BH}$ relation. In fig. 7 we indicate by an arrow how the velocity dispersion estimate of luminous objects would move
if the best fit to the F-J relation ($L\sim \sigma^{3.3}$) for the Seyferts (dotted line in fig 6) were used.

\begin{inlinefigure}
\centerline{\includegraphics[width=1.2\linewidth]{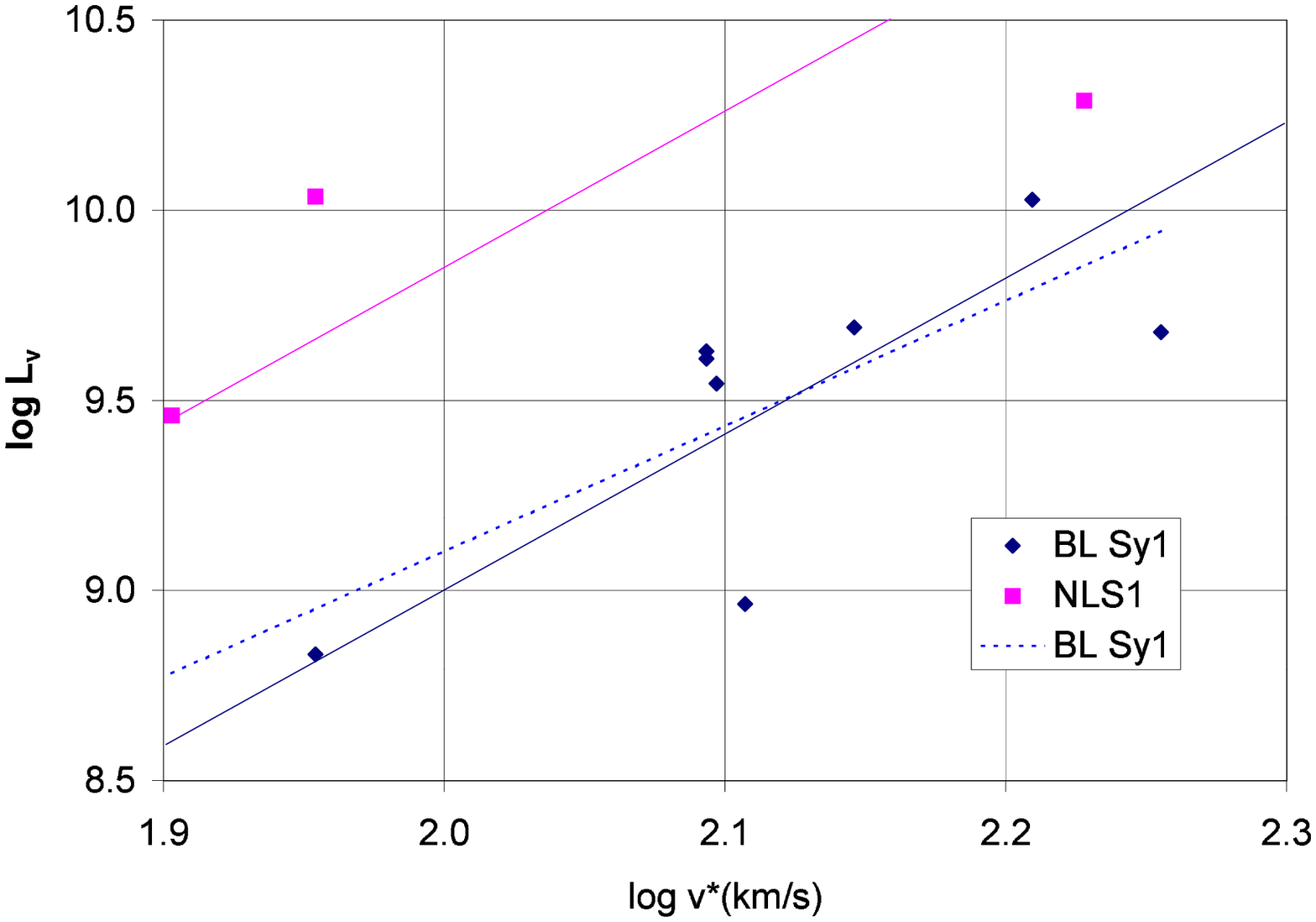}}
\figcaption{Bulge luminosity plotted against stellar velocity dispersion for Seyfert galaxies.
Blue 
Triangles indicate broad-line Seyfert 1s, pink 
squares denote NLS1s.
The solid lines denote the Faber-Jackson relation ($L\sim \sigma^4$) normalized
for the broad and narrow line Seyferts, respectively, while the dashed line is the best fit to the broad line Seyferts.}
\end{inlinefigure}

 In agreement with previous results (Gebhardt 2000b, Merritt \& Ferrarese 2001b) the 
NLANs (solid and open pink squares in fig 7) are more  or less consistent with the  
$M_{\rm BH}-\sigma$ relation, although they do lie at the lower end of the distribution.

The agreement of the $M_{\rm BH}-\sigma$ relations of inactive galaxies and AGNs (broad and narrow line)
suggests  that the virial masses are essentially correct also for NLANs.
The tendency of the NLANs to have low $M_{\rm BH}/\sigma$ ratios suggests that their lower BH-bulge
ratios go along with relatively lower BH masses than broad line AGNs.
%
In order to settle this issue, more stellar 
velocity-dispersion measurements of AGNs are required, in particular of NLS1s.


\begin{inlinefigure}
\centerline{\includegraphics[width=1.2\linewidth]{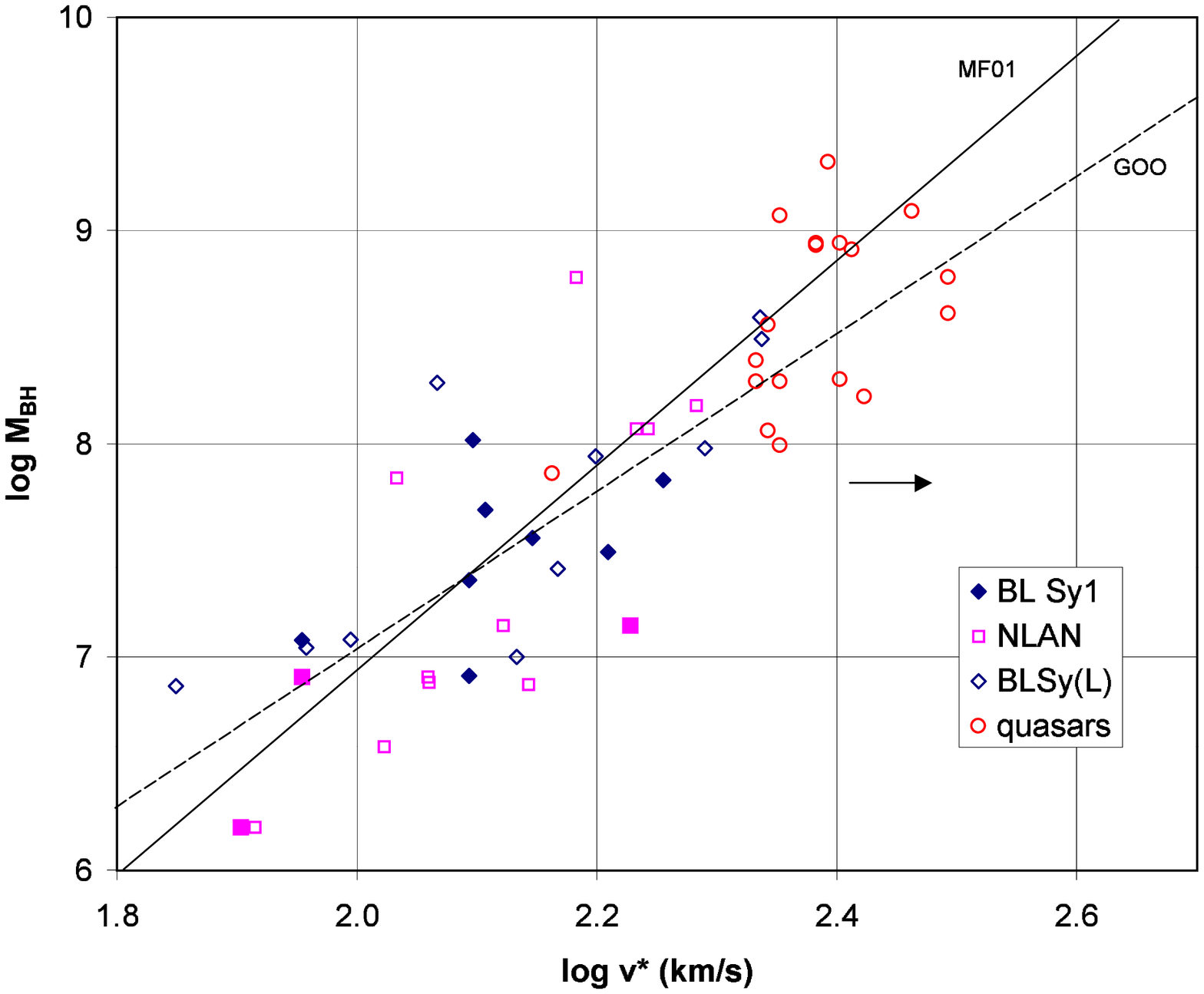}}
\figcaption{Black hole mass of AGNs plotted against the stellar velocity dispersion ($v^*=\sigma$).
Blue triangles are broad line Seyferts,  pink 
squares denote NLS1s and red 
circles denote quasars.
Solid symbols dennote Seyferts with measured $\sigma$, open ones denote Seyferts 
for which $\sigma$ has been estimated from the Faber-Jackson relation (see text).
The dashed and solid lines show the $\sigma-M_{\rm BH}$ relation of galaxies 
(Gebhardt  \et 2000a and Ferrarese \& Merritt 2000, respectively). 
The arrow shows how the velocity 
dispersion estimate of luminous objects would typically move
if the best fit to the F-J relation for the Seyferts (fig 6) were used.
}
\end{inlinefigure}

The difference between NLANs and broad line AGNs in  
the BH-bulge relation is much larger than the difference in the BH-velocity-dispersion relation. 
This may be related to the bulge mass 
(and luminosity) being integral, extended properties, while the stellar velocity dispersion 
is more closely related to the inner part of the bulge.
There is more evidence that the \bh mass in AGN is correlated with the velocity dispersion (and hence the 
virial mass) on the 
scale of the central part of the bulge -  the narrow line region. This is supported by
the correlation between 
the \bh mass and the velocity dispersion of the narrow emission-line gas, as measured by the [OIII] 
lines in Seyfert 1 nuclei (Wandel and Mushotzky 1986; Nelson 2000).

\section{Do NLANs really have lower BH/bulge ratios?}

We have seen that broad line AGNs follow a similar BH-bulge relation as inactive galaxies.
The question remains what can be the cause of the lower BH/bulge ratio in the narrow-line 
AGNs, and whether this effect is real or apparent.

Several effects have been mentioned that could cause such an apparent lower BH/bulge ratio 
(e.g. Wandel 2000):
(i) the \rev - virial method may
systematically underestimate the BH mass in NLANs, 
(ii) the measured bulge luminosity in NLANs may be 
systematically too large.
Alternatively, the difference between the two groups could be real:
either (iii) NLANs do have intrinsically larger (or brighter) bulges, or
(iv) they have systematically smaller \bhs. 
In section 5 we have seen that effects (ii) and (iii) 
are supported by the $\sigma-M_{\rm BH}$ relation, as
the narrow-line Seyferts with measured velocity 
dispersion (MRK 110, MRK 590 and NGC 4051) 
seem to have larger bulge 
luminosity than broad line Seyferts with comparable 
velocity dispersions (fig. 6). 
On the other hand, also (i) and (iv) cannot 
be ruled out, as the \rev virial BH masses of 
MRK 590 and NGC 4051 are by a factor 2-3 lower than the average $M_{\rm BH}-\sigma$ relation 
of broad line Seyferts and inactive galaxies (fig.7).

There are a number of observational errors 
that may cause an apparent difference in the
\bh - bulge luminosity relation between NLANs and  classic AGNs and inactive 
galaxies, which we discuss below.

\subsection{Errors in estimating bulge luminosity}

As suggested above, the lower $M_{\rm BH}-L_{\rm bulge}$ values of NLAN
could be accounted for if their bulge luminosity 
are systematically over estimated.
As we saw, this was indeed the case when the Simien-deVaucouleurs method was used for Seyfert galaxies. 
Could a similar problem exist in the NLAN set?
Two of the 3 NLS1s with measured velocity dispersion have low $M_{\rm BH}/\sigma$ 
values which may indicate they have relatively small \bhs . 
However, the location of the NLANs in the $M_{\rm BH}-\sigma$ plot (fig 7) 
with respect to the inactive and broad
line AGNs indicates that the BH mass can account for only a small part of
the difference, suggesting that NLS1s tend to have relatively large bulges.
It is however difficult to see why the bulge magnitude of NLANs be
{\it systematically} over-estimated by a factor of $\sim$10 (required to explain
the difference in the BH/bulge ratio) compared with the 
broad line AGNs in the sample.

\subsection{Bias introduced by the stellar kinematics method}

Is it possible that NLANs represent a larger 
population of galaxies with low \bh mass to bulge luminosity ratios, which is under-represented in
the presently available MBH sample? This could be the case for MBHs detected 
using stellar 
dynamics, because this method cannot detect small \bhs, and the detection-limit 
increases with distance. In a resolution-limited method, this would infer a 
 lower detection limit which increases with luminosity.
This hypothesis is supported by the distribution of dynamically estimated \bh masses in quiescent galaxies.
There are only four inactive galaxies with BH masses under 
~$ 10^7\ms$:  the Milky Way, M32 
and NGC 7457 detected using stellar dynamics and NGC4945 - with maser dynamics. The two latter galaxies
also have low \bh -to bulge luminosity ratios, comparable to the NLAN average 
ratio.
In angular-resolution limited methods, the MBH detection limit 
is correlated with bulge luminosity: for more luminous
bulges the detection limit is higher, because the stellar velocity
dispersion is higher (the Faber-Jackson relation). In order to detect the 
dynamic effect of a MBH it is necessary to observe closer to the center,
while the most luminous galaxies tend to be at  larger distances, so for
a given angular resolution the MBH detection limit is higher.
This may imply that the sample is biased towards larger
MBHs, as present stellar-dynamical methods are ineffective
for detecting MBHs below $\sim 10^7 \ms$ (except in the nearest galaxies).
Since the BLR method is not subject to this constraint,
NLANs may represent a
low-mass BH population.

\subsection{The \bh influence on the stellar velocity dispersion}

In section 5 we have seen that the NLANs are not significantly outlying the $\sigma-M_{\rm BH}$ relation,
while they are located far below the BH-bulge relation.
In principle this could be caused by the BH influence on the measured
velocity dispersion.
In NLANs the region influenced by the BH would be relatively smaller 
(either because the BH has a lower mass or because the bulge is larger, and 
hence has a larger velocity dispersion).
This can reduce the difference between narrow and broad line AGNs when $\sigma$ 
rather than $L_{\rm bulge}$ is considered. However, as we show below, the BH influence turns out to be negligible for almost all AGNs in our sample.
The effect of the massive BH on the stellar velocity dispersion can be 
estimated by comparing the expected
velocity field $(v\propto (M_{\rm BH}/r)^{1/2}$) due to the MBH
and the velocity dispersion of the host bulge.
The \bh enhances the velocity dispersion in an observed region 
if its mass is bigger or comparable to the stellar mass
within the radius corresponding to the projected angular size.
This can be measured by the dimensionless quantity
$$m = GM_{\rm BH}/\sigma^2 \theta D,$$
where $\sigma$ is the stellar velocity
dispersion in the bulge of the host galaxy,
$\theta$ is the angular size (e.g. the width of the slit) and
$D$ is the distance to the galaxy.
In other words, $m$ is roughly the ratio of the BH mass to the stellar bulge
mass inside the radius corresponding to the angular size being sampled.
The nuclear velocity dispersion is typically sampled 
with a slit width of 1-2\arcsec (Ferrarese \et 2001 use a slit of 2x4\arcsec ).
Since the measured velocity dispersion is weighted by the 
brightness along the line-of-sight, and since the brightness-density increases 
steeply at the center, the effective value of $\theta$ is probably smaller than the slit width.
The BH influence parameter can be written as
$m= 0.9 (M/10^8 M_\odot )\sigma_2^{-2}(D/10 Mpc)^{-1}$,
where $\theta$ has been assumed to be $1\arcsec$.
We have calculated $m$ for the 11 Seyferts with velocity dispersion measurements. 
The three NLS1s have $m\sim 0.01$, lower than the BL Seyferts
which have typically $m\sim 0.03-0.1$, with the exception of NGC 3227 which has
$m=0.24$. We conclude that the BH influence on the velocity dispersion in the observed AGNs is negligible. 

\subsection{Inclination and BLR geometry}

If the broad emission-line region (BLR) has a flattened geometry and the distribution
of inclination angles is biased towards face-on inclinations 
(as one might expect
in the unified model) the velocity inferred from the observed 
line width may be smaller than the 3D velocity dispersion, 
depending on the 
inclination and the amount of flattening. 
For example, for a flat BLR configuration viewed 
at an inclination angle $i$ 
(where $i=90$ corresponds to face-on), the line-of-sight 
velocity is smaller than the 3-D velocity by a factor of sin($i$). 
The inferred mass would thus be smaller 
than the actual mass by a factor of 
$\sin^2(i)$. For a 
random distribution of inclination angles, 
the average inferred mass would be decreased 
by this factor weighted by the distribution, 
$<\sin^2(i)>=4\pi\int_0^{\pi/2} \sin^2(i)\cos(i){\rm d}i/4\pi=1/3$.
Actually, the distribution is probably 
not random. In the unified scheme Seyfert 1 nuclei are viewed more 
face-on, within an opening angle of $\la 60$ degrees, 
so that the weighted
average mass-reduction factor is 
$<\sin^2(i)>=4\pi\int_0^{\pi/6} \sin^2(i)\cos(i){\rm d}i/2\pi=1/12$.
If the BLR distribution 
is not flat but with an angular distribution of orbits spanning an angle 
$\delta$, the effect 
would be less - between sin($i$) and
sin$(\delta+i)$, depending on the distribution of orbits, so that roughly
$<\sin^2(i+\delta)>~\ga 0.1$. 

This scenario can explain the lower $M_{\rm BH}/L_{\rm bulge}$ values of NLANs.
The distribution of inclination angles can also explain the larger scatter of NLANs, 
compared with broad line Seyferts.
In this scenario, NLANs are AGNs with a flattened
BLR geometry. seen nearly face-on. 

\subsection{Uncertainties and non-virial BLR dynamics}

Is it possible that the \bh masses of NLANs are systematically 
underestimated by the \rev - virial method? 
The \rev masses of all AGN seem to be more or less correct, in light of  
the good agreement between AGNs and quiescent galaxies 
in the case of the \bh - velocity dispersion relation.
The uncertainty in the \rev virial method is not well known. While for 
individual objects 
a factor of 2-3 may be representative, the sample average error is probably 
much smaller. This comes on top of the measurement uncertainty; the
uncertainty in the BLR size (calculated from the cross correlation of the continuum and line
light curves) can be quite large for
objects with poor sampling or low variability.
Note  that if the virial assumption is incorrect and the gas in the line emitting gas 
in NLANs is unbound, the gas velocity is actually larger
than Keplerian, the \bh mass would have been {\it overestimated}, which 
would increase the BH-bulge
discrepancy between NLANs and broad line Seyfert galaxies (and
inactive galaxies). Similarly, if the BH mass of 
broad line AGNs were underestimated,
their present good agreement with the BH-bulge relation 
of quiescent galaxies would deteriorate, as demonstrated in section 3.4.

More generally, we may ask how reliable are the BH masses 
derived by the \rev virial method?
Could there be a systematic effect?
The best resolution of HST in relatively nearby galaxies translates into
a distance of a few tens of parsecs from the MBH. Reverberation mapping 
of the broad emission line region in
AGNs gives a much closer view - a few light days from the center. 
Assuming the line-emitting matter is gravitationally bound, 
having a Keplerian velocity dispersion , 
it is possible to estimate the virial mass:
$M\approx G^{-1}rv^2 .$
 This expression may be approximately valid also in the case
the line emitting gas is not bound,
such as radiation-driven motions and disk-wind models (e.g. Murray \et 1998).
The main problem in estimating the virial mass from the \el data is to obtain
a reliable estimate of the size of the BLR, and to correctly identify the
line width with the velocity dispersion in the gas (WPM, Krolik 2001). 
WPM used the continuum/emission-line cross-correlation function to measure the
responsivity-weighted radius $c\tau$ of the BLR (Koratkar \& Gaskell 1991), 
and the variable (rms) component of the spectrum to measure the velocity dispersion in the same
part of the gas, which is used to calculate the BLR size, automatically 
excluding constant features such as narrow emission lines and 
galactic absorption. Kaspi \et (2000) find that for most objects in their sample, 
similar mass estimates are obtained also using either the rms as or the mean FWHM.  
The case for a central MBH dominating the kinematics in the broad emission-line region is supported 
by the Keplerian velocity profile ($v \propto r^{-1/2}$) detected in NGC 5548 
(Peterson \& Wandel 1999) and NCG 7469 and 3C390.3 (Peterson \& Wandel 2000). 
Although different \els have quite different line widths,
the delays also vary from line to line in such a way that
the virial masses derived from different emission lines are 
all consistent with a single value.
This demonstrates the case for a
Keplerian velocity dispersion in the line-width/time-delay data.
Another support for the validity of the \rev virial \bh mass estimates comes from the 
consistency of the \bh-bulge relations for  AGN and inactive galaxies: the $M_{\rm BH}-\sigma$ 
relation (Ferrarese \et 2001), and the $M_{\rm BH}-L_{\rm bulge}$ relation (this work).

\subsection{Evolution}

The different BH-bulge 
relation of NLANs compared with broad-line AGNs and inactive galaxies 
could suggest that 
NLANs may be intrinsically different. One possibility is that they are in a different 
evolutionary stage.
 Wandel (1999) suggested that in AGNs which are radiating near the
Eddington luminosity, the central \bh 
accretes within a relatively short time scale most of the readily available 
matter. BHs accreting near the \ed limit
are still growing, hence their mass (relative to the bulge) may be smaller
than in AGNs with lower $L/M$ ratios such as Seyferts 1s which may be past 
the most active phase. Similarly, MBHs in normal galaxies
may be the remnants of a past luminous AGN phase, and their BHs did not
grow significantly after the decline in the accretion rate. This 
could explain 
the discrepancy between narrow and broad line AGNs 
(NLANs tend to be near the \ed limit), 
as well as the similar BH-bulge relations of broad line AGNs and inactive galaxies.
This scenario has been followed with a simple model 
calculation  by Wang, Biermann and Wandel (2000) who 
examined the relative contribution of mergers to the bulge and BH mass. For 
reasonable parameters they find that a BH/bulge mass ratio of $\sim0.002$ 
is obtained by evolution of the MBH 
due to Eddington-limited accretion enhanced due to mergers.

\section{Summary}
We re-examine the purported discrepancy between the \bh - bulge luminosity relation of 
MBHs in nearby galaxies (measured with stellar- and gas dynamics methods) and \rev -mapped 
Seyfert nuclei (Wandel 1999). Using
updated data. we find that the BH/bulge relation of AGNs and inactive galaxies 
are in good agreement over 3 orders of magnitude in bulge luminosity or BH mass.
We show that the apparent difference reported previously originated from 
overestimated BH masses in quiescent galaxies, overestimated bulge luminosity
for the Seyferts and including narrow-line Seyfert 1s which seem to have genuinely low BH/bulge ratios. 
We find that the lower BH/bulge ratio of narrow line AGNs is part of a more general inverse 
correlation of the BH/bulge ratio with the emission-line width. 
Considering the BH-velocity dispersion correlation, the lower BH/bulge ratio of narrow-line AGNs
seems to be related (at least in part) to the 
host, as narrow and broad line AGNs 
(as well as inactive galaxies) seem to have  
a similar $\sigma -M_{\rm BH}$ relation. 
We predict the velocity dispersion of high luminosity AGN using the Faber Jackson relation derived from Seyfert galaxies for with a measured $\sigma$.
We discuss three classes of explanations for the lower BH/bulge ratios of narrow-line AGNs:
\bh mass - bulge luminosity relation of Seyfert galaxies and MBHs in normal galaxies:
Observational or method-related errors or bias, intrinsic and
orientation-related effects and Evolution.

\acknowledgments
I acknowledge support from BSF grant  1999-336 and the hospitality of the astronomy 
department at UCLA, as well as helpful discussions with Laura Ferrarese, Karl Gebhardt, Luis Ho, Ari Laor and David Merritt.

\end{document}